\documentclass[3p,twocolumn]{elsarticle}

\usepackage{amsfonts,amsmath,amssymb}
\usepackage{graphicx,color}

\def\blfootnote{\xdef\@thefnmark{}\@footnotetext}

\begin{document}

\begin{frontmatter}

\title{Charged-particle multiplicity dependence of charm-baryon-to-meson ratio in high-energy proton-proton collisions}

\author[a]{Yu Chen}
\author[a]{Min He}

\address[a]{Department of Applied Physics, Nanjing University of Science and Technology, Nanjing 210094, China}

\date{\today}

\begin{abstract}
We propose that the charged-particle multiplicity dependence of the charm-baryon-to-meson ratio observed in high-energy $pp$ collisions can be explained by canonical treatment of quantum charges in the statistical hadronization model (SHM). Taking the full particle listings of PDG complemented by additional charm-baryon states from relativistic quark model predictions, we evaluate the canonical partition function and the charm-hadron chemical factors that measure the canonical suppression arising from the requirement of strict conservation of quantum charges. We demonstrate that, while charm number conservation induces common suppression on the production of both charm-baryons and -mesons, baryon (strangeness) number conservation causes further suppression on charm-baryons (charm-strange mesons) relative to nonstrange charm-mesons, thereby resulting in a decreasing $\Lambda_c/D^0$ ($D_s/D^0$) ratio toward smaller multiplicity events. The charm-hadron thermal densities thus computed are then used as pertinent weights to perform charm-quark fragmentation simulations yielding $p_T$-dependent $\Lambda_c/D^0$ and $D_s/D^0$ ratios at varying multiplicities in fair agreement with ALICE measurements.
\end{abstract}

\begin{keyword}
Heavy flavor production \sep Charm baryons \sep Canonical ensemble \sep Proton-proton collisions
\end{keyword}

\end{frontmatter}

\section{Introduction
\label{sec_intro}}
Factorization provides an approach to compute the production of heavy-flavor (HF) hadrons in high-energy hadronic collisions within the framework of perturbative QCD. This approach involves the convolution of the parton distribution functions (PDFs) of the colliding hadrons, the hard-scattering cross sections at the partonic level and the fragmentation functions (FFs). While the PDFs are considered to be universal and usually taken from global fits of various sets of empirical data~\cite{Placakyte:2011az}, and the partonic cross sections are calculated as perturbative series in powers of the strong-coupling constant, the FFs modelling the nonperturbative hadronization processes represent a major source of theoretical uncertainty.

FFs parameterize the percentage of the quark energy transferred to the produced hadron and the fraction (weight) of a heavy quark hadronizing into a particular hadron species. Under the assumption that they are universal across different collision systems, FFs are usually tuned against electron-positron ($e^+e^-$) collision data~\cite{Lisovyi:2015uqa}. However, this assumption of universality has been challenged by the measurement of production of heavy baryons, which is in particular sensitive to the fragmentation mechanism of heavy quarks (HQs)~\cite{Acharya:2017kfy}. For example, in the case of bottom quarks, the $b\rightarrow\Lambda_b^0$ fragmentation fraction was reported not to be compatible with that deduced from LEP data~\cite{Amhis:2016xyh}. In the case of charm quarks, fragmentation calculations based on fit of the $e^+e^-$ knowledge in the $k_T$-factorization approach~\cite{Maciula:2018iuh} or the general-mass variable-flavor number scheme~\cite{Kniehl:2020szu} reported a substantial deficiency in the $\Lambda_c$ production cross section as compared to the measurement by the ALICE collaboration at mid-rapidity in $\sqrt{s}=7$\,TeV $pp$ collisions~\cite{Acharya:2017kfy}; and therefore the the pertinent $\Lambda_c/D^0\simeq0.54$ ratio, which is also regarded as a sensitive measure of novel recombination hadronization mechanisms in heavy-ion collisions~\cite{Acharya:2018ckj,Sirunyan:2019fnc,He:2019vgs}, was much underestiamted~\cite{Maciula:2018iuh,Kniehl:2020szu}. String fragmentation invoking color reconnection mechanisms accountes for multi-parton interactions in high-energy $pp$ collisions and turns out to enhance the baryon production in both strange and charm sector~\cite{Bierlich:2015rha}.

Another way of determining the fragmentation fractions of a heavy quark hadronizing into various kinds of HF hadrons is given by their thermal weights at a hadronization ``temperature" $T_H$ in the statistical hadronization model (SHM)~\cite{BraunMunzinger:2003zd,Andronic:2017pug}, given that {\it relative} chemical equilibrium among them may be reached in the quark-rich environment such as high-energy $pp$ collisions. It was shown~\cite{He:2019tik} that by incorporating a large set of hitherto unobserved charm-baryon states as inspired by relativistic quark model (RQM)~\cite{Ebert:2011kk} computations into the grand-canonical SHM~\cite{Kuznetsova:2006bh,Andronic:2007zu}, the large $\Lambda_c/D^0\simeq0.54$ reported by ALICE~\cite{Acharya:2017kfy} could be well accounted for, simply augmented by the enhanced feedowns from those ``missing" charm-baryons in particular in the low momentum regime.

While the aforementioned $\Lambda_c/D^0\simeq0.54$ was measured in minimum bias $pp$ collisions, measurement of the dependence of HF production in such collisions on the charged particle multiplicity ($dN_{\rm ch}/d\eta$) may provide insight into the multi-parton interactions as well as the interplay between the hard and soft mechanisms in particle production~\cite{Adam:2015ota,Acharya:2020pit,Sirunyan:2020zzb,Aaij:2020hpf,Deb:2020ige}. Indeed, study of $pp$ collisions at varying $dN_{\rm ch}/d\eta$ is of considerable interest as phenomena typically indicative of the formation of a deconfined quark-gluon medium have been observed in high-multiplicity $pp$ collisions, such as the long-range two-particle azimuthal correlations on the near side~\cite{Khachatryan:2016txc} and the enhanced production of multi-strange hadrons~\cite{ALICE:2017jyt}, opening up an angle for examining the final state effects known from heavy-ion collisions. As far as the $\Lambda_c/D^0$ is concerned, its dependence on $dN_{\rm ch}/d\eta$~\cite{CHill:HP2020} would reflect the modification of the fragmentation mechanisms of charm quarks as the surrounding environment changes.

In the present work, we generalize the grand-canonical SHM investigation~\cite{He:2019tik} of charm-baryon production to the case of canonical SHM, and explore in how far the observed multiplicity dependence of $\Lambda_c/D^0$ in high-energy $pp$ collisions~\cite{CHill:HP2020} can be explained by the additional canonical suppression of the production of charm-baryons relative to charm-mesons. Particles entering the evaluation of the canonical partition function are taken from the full listings of PDG~\cite{Tanabashi:2018oca} (excluding bottom particles), with charm-baryons still augmented by the additional states predicted by the relativistic quark model (RQM) computations~\cite{He:2019tik,Ebert:2011kk}. The existence of those ``missing" charm-baryon states is supported by the direct measurement of charm-baryon spectroscopy on lattice~\cite{Padmanath:2014bxa} as well on the analysis of the partial pressure of open-charm states and charm-quark susceptibilities in thermal lattice QCD~\cite{Bazavov:2014yba}. We calculate the chemical factors of each charm-hadron as a measure of canonical suppression relative to the grand-canonical (large volume) limit. We demonstrate that, while strict charm number conservation as imposed by canonical treatment of SHM induces common suppression on both charm-baryons and -mesons, requirement of strict baryon number conservation causes further suppression on charm-baryons, resulting in a decreasing $\Lambda_c/D^0$ toward smaller system size (or multiplicity $dN_{\rm ch}/d\eta$). Similarly, strict strangeness conservation causes further suppression on charm-strange mesons and leads to a multiplicity-dependent $D_s/D^0$ ratio. Using thus computed thermal densities of each charm-hadron as pertinent weights, we conduct a simulation of the charm quark fragmentation within the FONLL scheme~\cite{Cacciari:1998it,Cacciari:2012ny} and compute the transverse momentum ($p_T$) dependent $\Lambda_c/D^0$ and $D_s/D^0$ at different $dN_{\rm ch}/d\eta$.

\section{Canonical ensemble SHM}
\label{sec_canonicalshm}
\subsection{Canonical partition function}
\label{ssec_canonical-Z}
When applied to heavy-ion collisions, SHM is usually implemented in a grand-canonical ensemble (GCE) of an ideal hadron resonance gas (HRG), in which quantum charges, such as electric charge $Q$, baryon number $B$, strangeness $S$ and charm number $C$ are conserved {\it on average} and regulated by the corresponding chemical potentials $\vec \mu=(\mu_Q,\mu_B,\mu_S,\mu_C)$. For hadrons whose mass is much larger than the typical hadronization temperature $T_H$ and thus the quantum statistics effect can be safely neglected, the primary mean number of the $j$-th hadron produced from this GCE-SHM is then given by
\begin{equation}\label{GCE_mean_number}
\langle N_j\rangle^{GCE}=\gamma_s^{N_{sj}}\gamma_c^{N_{cj}}z_je^{\vec{\mu}\cdot\vec{q_j}/T_H},
\end{equation}
where $\vec q_j=(Q_j,B_j,S_j,C_j)$ is the corresponding quantum charges and $z_j$ is the one-particle partition function of the $j$-th hadron of mass $m_j$ and spin $J_j$
\begin{align}
z_j=(2J_j+1)\frac{VT_H}{2\pi^2}m_j^2K_2(\frac{m_j}{T_H}),
\end{align}
which is just the chemical-equilibrium multiplicity of the $j$-th hadron in the grand-canonical limit under the Boltzmann approximation, with $K_2$ being the modified Bessel function of the second order. In Eq.~(\ref{GCE_mean_number}), $\gamma_s$ and $\gamma_c$ are fugacities to account for the deviation of strange and charm production from chemical equilibrium, respectively, for hadrons containing $N_{sj}$ ($N_{cj}$) strange (charm) valence quarks and antiquarks. As the hadronization temperature $T_H$ is usually assumed to be a universal parameter in the GCE implementation of SHM, the hadron yield ratios would be the same in all collision centralities.

However, it has long been pointed out~\cite{Rafelski:1980gk,Hagedorn:1984uy} that requirement of {\it exact} conservation of quantum charges as implemented in the canonical ensemble (CE) treatment of SHM becomes increasingly important for sufficiently small reaction volumes like those in elementary $pp$ and $e^+e^-$ collisions; thereby a system size (or charged particle multiplicity $dN_{ch}/d\eta$) dependence of hadron production would follow from the canonical corrections~\cite{Vovchenko:2019kes}. For a collision system having conserved quantum charges $\vec Q=(Q,B,S,C)$, the CE-SHM partition function at a given hadronization temperature $T_H$ reads~\cite{Becattini:1995if,Becattini:1997rv,Becattini:2009sc}
\begin{align}\label{CE_partition_function}
Z(\vec Q)=\frac{1}{(2\pi)^4}\int_0^{2\pi}d^4\phi e^{i\vec Q\cdot\vec\phi}{\rm exp}[\sum_j \gamma_s^{N_{sj}} \gamma_c^{N_{cj}} e^{-i\vec q_j\cdot\vec\phi}z_j],
\end{align}
where the summation $\sum_j$ is over {\it all} hadrons and the volume in $z_j$ now should be understood as the correlation volume in which the quantum charges are strictly conserved (it will be still denoted as $V$ throughout the work, except in Sec.~\ref{sec_frag} it is replaced with $V_c$). Here, again the Boltzmann approximation is assumed, which causes an error of few percent for pions (for typical hadronization temperature
$T_H=160-170$\,MeV) when resonance decays are taken into account, and much less for all other hadrons~\cite{Becattini:1997rv}. Note that, unlike the case of GCE-SHM, the logarithm of the canonical partition function ${\rm log}Z(\vec Q)$ does not scale linearly with the volume.

The primary mean number of the $j$-th hadron produced from this CE-SHM is then given by~\cite{Becattini:1997rv,Becattini:2009sc}
\begin{align}\label{CE_mean_number}
\langle N_j\rangle^{CE}=\gamma_s^{N_{sj}}\gamma_c^{N_{cj}}z_j\frac{Z(\vec Q-\vec q_j)}{Z(\vec Q)}.
\end{align}
Comparing Eq.~(\ref{GCE_mean_number}) and Eq.~(\ref{CE_mean_number}) reveals that the chemical potential term in the GCE-SHM associated with conservation of quantum charges {\it on average} is now replaced with the {\it chemical factor} $Z(\vec Q-\vec q_j)/Z(\vec Q)$ in the CE-SHM arising from the requirement of {\it exact} conservation of quantum charges. It will be shown (see Sec.~\ref{sec_canonicalsuppression}) that the chemical factors for charged hadrons ({\it i.e.}, $\vec q_j\neq 0$) are always less than unity in a completely neutral system ({\it i.e.}, $\vec Q=0$ and thus $\vec\mu=0$) and reach their grand canonical value of unity at asymptotically large volumes, therefore characterizing the {\it canonical suppression} of the production of charged hadrons at finite volume.

\subsection{Evaluation of the canonical partition function}
\label{ssec_evaluating-Z}

To evaluate the canonical partition function, one divides hadrons to be summed over in the exponential of Eq.(\ref{CE_partition_function}) into three categories~\cite{Keranen:2001pr}. The first category consists of completely neutral mesons ($M^0$) with $\vec q_j=(Q_j,B_j,S_j,C_j)=(0,0,0,0)$, {\it e.g.}, $\pi^0$, $\eta$, $\rho^0$, $a_0(980)^0$, $\phi$, $J/\psi$. Clearly their contribution to the canonical partition function can be factorized out, yielding a multiplicative term
\begin{equation}
Z_0={\rm exp}[\sum_{j=M^0}\gamma_s^{N_{sj}}\gamma_c^{N_{cj}}z_j]=\prod_{j=M^0}{\rm exp}(\gamma_s^{N_{sj}}\gamma_c^{N_{cj}}z_j).
\end{equation}

The second category refers to charged mesons, including positively charged mesons ($M^+$) with at least one of $Q_j=+1$, $S_j=+1$ or $C_j=+1$ (but definitely $B_j=0$), {\it e.g.}, $\pi^+$, $\rho^+$, $K^+$, $K^0$, $K^{*+}$, $K^{*0}$, $D^0$, $D^+$, $D^{*0}$, $D^{*+}$, $D_s^+$, $D_s^{*+}$, and their anti-mesons ($M^-$), {\it e.g.}, $\pi^-$, $\rho^-$, $K^-$, $\bar{K}^0$, $K^{*-}$, $\bar{K}^{*0}$, $\bar{D}^0$, $D^-$, $\bar{D}^{*0}$, $D^{*-}$, $D_s^-$, $D_s^{*-}$. Noting that $M^+$'s and $M^-$'s always have opposite charges but the same $N_{sj}$ and $N_{cj}$, the contribution of the charged mesons to the exponential in Eq.~(\ref{CE_partition_function})
\begin{align}
&{\rm exp}[\sum_{j=M^+,M^-}\gamma_s^{N_{sj}}\gamma_c^{N_{cj}}z_je^{-i(Q_j\phi_Q+S_j\phi_S+C_j\phi_C)}]\nonumber\\
&={\rm exp}[\sum_{j=M^+}\gamma_s^{N_{sj}}\gamma_c^{N_{cj}}z_je^{-i(Q_j\phi_Q+S_j\phi_S+C_j\phi_C)}\nonumber\\
&+\sum_{j=M^+}\gamma_s^{N_{sj}}\gamma_c^{N_{cj}}z_je^{i(Q_j\phi_Q+S_j\phi_S+C_j\phi_C)}]\nonumber\\
&={\rm exp}[2\sum_{j=M^+}\gamma_s^{N_{sj}}\gamma_c^{N_{cj}}z_j{\rm cos}(Q_j\phi_Q+S_j\phi_S+C_j\phi_C)].
\end{align}

The third category comprises baryons ($B$) with $B_j=+1$ and antibaryons ($\bar{B}$) with $B_j=-1$. Obviously $B$'s and $\bar{B}$'s carry opposite values of all other charges ($Q_j$, $S_j$ and $C_j$), so that their contribution to the exponential in Eq.~(\ref{CE_partition_function})
\begin{align}\label{ZCE-mesons}
&{\rm exp}[\sum_{j=B,{\bar B}}\gamma_s^{N_{sj}}\gamma_c^{N_{cj}}z_je^{-i(Q_j\phi_Q+B_j\phi_B+S_j\phi_S+C_j\phi_C)}]\nonumber\\
&={\rm exp}[e^{-i\phi_B}\sum_{j=B}\gamma_s^{N_{sj}}\gamma_c^{N_{cj}}z_je^{-i(Q_j\phi_Q+S_j\phi_S+C_j\phi_C)}\nonumber\\
&+e^{i\phi_B}\sum_{j=B}\gamma_s^{N_{sj}}\gamma_c^{N_{cj}}z_je^{i(Q_j\phi_Q+S_j\phi_S+C_j\phi_C)}\nonumber\\
&={\rm exp}[e^{-i\phi_B}\omega+e^{i\phi_B}\omega^*]\nonumber\\
&={\rm exp}[2|\omega|{\rm cos}(\phi_B-{\rm arg}\omega)],
\end{align}
where
\begin{align}
\omega&=\omega(\phi_Q,\phi_S,\phi_C)=|\omega|e^{i{\rm arg}\omega}\nonumber\\
&=\sum_{j=B}\gamma_s^{N_{sj}}\gamma_c^{N_{cj}}z_je^{-i(Q_j\phi_Q+S_j\phi_S+C_j\phi_C)}.
\end{align}

Now combining contributions from all these three categories gives rise to the canonical partition function
\begin{align}
&Z(\vec Q)=\frac{Z_0}{(2\pi)^3}\int_0^{2\pi}d^3\phi {\cos}(Q\phi_Q+S\phi_S+C\phi_C+B{\rm arg}\omega)\nonumber\\
&\times I_B(2|\omega|)\times e^{2\sum_{j=M^+}\gamma_s^{N_{sj}}\gamma_c^{N_{cj}}z_j{\rm cos}(Q_j\phi_Q+S_j\phi_S+C_j\phi_C)},
\end{align}
where $I_B(2|\omega|)=\frac{1}{2\pi}\int_0^{2\pi}d\phi_Be^{2|\omega|{\rm cos}\phi_B}{\rm cos}(B\phi_B)$ is the modified Bessel function
of the first kind and order $B$.

\begin{figure} [!tb]
\vspace{-0.5cm}
\includegraphics[width=1.05\columnwidth]{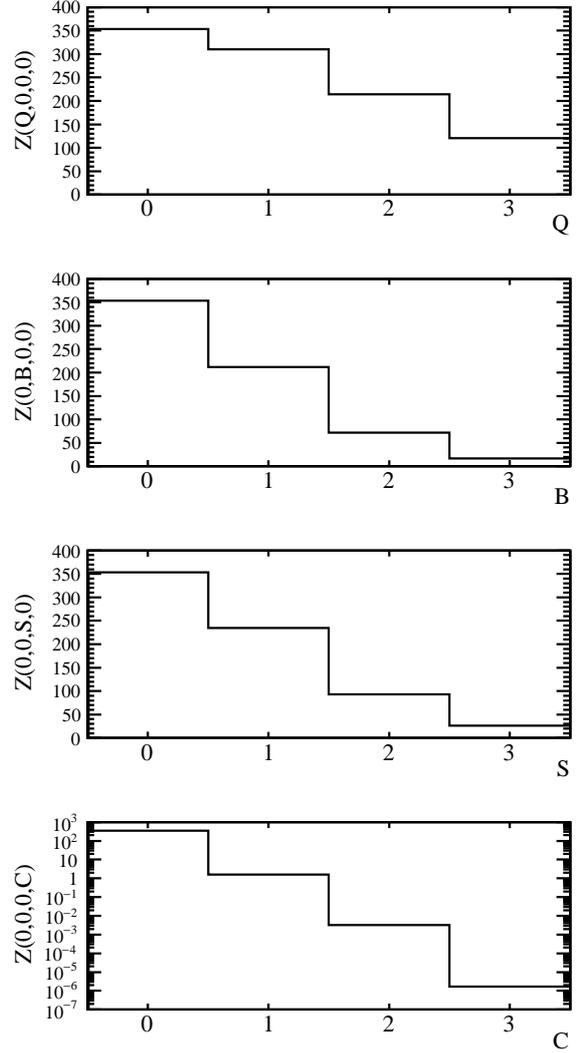}
\vspace{-0.5cm}
\caption{The behavior of the canonical partition function at $V=20~{\rm fm^3}$, $T_H=170$\,MeV and $\gamma_s=\gamma_c=1$, as a function of integer electric charge ($Q$), baryon number ($B$), strangeness ($S$) and charm number ($C$), respectively, with remaining quantum charges set to zero.}
\label{fig_Z(QBSC)}
\end{figure}

Taking the full particle listings in PDG~\cite{Tanabashi:2018oca} with charm-baryons augmented by additional states from relativistic quark model (RQM) computations~\cite{He:2019tik,Ebert:2011kk}, the canonical partition function is evaluated and the behavior as a function of increasing $Q, B, S, C$ at $V=20~{\rm fm^3}$, $T_H=170$\,MeV and $\gamma_s=\gamma_c=1$ is elucidated in Fig.~\ref{fig_Z(QBSC)}. The immediate observation is, while the dependence of
$Z(Q,0,0,0)$ on the electric charge is rather mild, $Z(0,B,0,0)$ and $Z(0,0,S,0)$ display a much stronger suppression as baryon number or strangeness increases. The suppression of the canonical partition function at finite quantum charge relative to $Z(0,0,0,0)$ can be understood as the chemical factor of a corresponding charged hadron in a completely neutral system. For instance, $Z(0,1,0,0)/Z(0,0,0,0)$ can be seen as the corresponding chemical factor of an antibaryon. Canonical treatment of SHM implies that, the creation of a baryon requires the accompanying creation of an antibaryon in order to fulfill exact conservation of baryon number, which is of course energy-expensive given that the mass of the lightest baryon ({\it i.e.}, the proton) $\sim 1$\,GeV is already several times the typical hadronization temperature $T_H$; in contrast, creation of an electrically-charged hadron can be compensated by generating an oppositely-electrically-charged pion to exactly conserve electric charge, which is less expensive in terms of energy and explains the milder dependence of $Z(Q,0,0,0)$. This argument generalizes to the case of $Z(0,0,0,C)$, where exact charm number conservation requires the simultaneous production of a very massive charm and anticharm pair, explaining the strongest decrease ($\sim$ two orders of magnitude, shown in a logarithmic scale in the lowest panel) of the $Z(0,0,0,C)$ upon each unit of increase of the charm number.

\section{Canonical suppression of charm-hadron multiplicities
\label{sec_canonicalsuppression}}

The ALICE measurement of $\Lambda_c/D^0$ in $\sqrt{s}=13$\,TeV $pp$ collisions was performed in the mid-rapidity region, where the matter created is nearly baryon-free~\cite{CHill:HP2020}. Hence we impose vanishing quantum charges $\vec Q=(Q,B,S,C)=(0,0,0,0)$ ({\it i.e.}, a completely neutral system) in the CE-SHM calculations in the following.

\subsection{Selective canonical suppression}
\label{ssec_selectivecanonical}

\begin{table}[!t]
\begin{tabular}{lcccccc}
\hline\noalign{\smallskip}
$hadron$      & CF & $\langle N_j\rangle^{CE}$ & $\langle N_j\rangle ^{GCE}$ \\
\noalign{\smallskip}\hline\noalign{\smallskip}
$\pi^0$       & 1.00000    &  1.13033 &  1.13033  \\
$\pi^+$       & 0.89365    &  1.00146 &  1.12064\\
$K^0$         & 0.80905    &  0.30430 &  0.37612\\
$K^+$         & 0.80961    &  0.30917 &  0.38187\\
$p$           & 0.66607    &  0.076786 &  0.11528 \\
$\Lambda$     & 0.66377    & 0.033375  &  0.050281 \\
$D^0$         & 0.00520    &  0.0000031128 &  0.00059834  \\
$D^+$         & 0.00518    &  0.0000030250 &  0.00058382    \\
$D_s^+$       & 0.00485    &  0.0000016972 &  0.00035016   \\
$\Lambda_c^+$ & 0.00458    &  0.0000006056 &  0.00013214\\
\noalign{\smallskip}\hline
\end{tabular}
\caption{Chemical factors (CF) and {\it direct} multiplicities of a selective set of light and charm hadrons at $V=20~{\rm fm^3}$, $T_H=170$\,MeV and $\gamma_s=\gamma_c=1$ in CE-SHM and in GCE-SHM, respectively}
\label{tab_CF-CEvsGCEmultiplicities}
\end{table}

In Table~\ref{tab_CF-CEvsGCEmultiplicities}, we show the chemical factors and {\it direct} ({\it i.e.}, without feeddowns) multiplicities $\langle N_j\rangle^{CE}$ and $\langle N_j\rangle^{GCE}$ of a selective set of light and charm hadrons calculated from CE-SHM and in GCE-SHM, respectively at $V=20~{\rm fm^3}$, $T_H=170$\,MeV and $\gamma_s=\gamma_c=1$. Clearly the chemical factor of a given hadron is obtained upon dividing $\langle N_j\rangle^{CE}$ by $\langle N_j\rangle^{GCE}$ in such a completely neutral system, which displays in general an increasing suppression for more massive charged hadrons. While the stronger canonical suppression of $\pi^+$ relative to $\pi^0$ and of $D^+$ relative to $D^0$ may be attributed to their carrying additional electric charge, further suppression of $K^+$ relative to $\pi^+$ and $D_s^+$ relative to $D^+$ is due to the requirement of strangeness conservation.

It is interesting to note that, when dividing out the common effects from canonical strangeness or charm suppression, the {\it relative} suppression ({\it i.e.}, the ratio of their chemical factors) of $\Lambda$ with respect to $K^0$ (denoted as $\Lambda/K^0$) $\simeq 0.820$ and of the $\Lambda_c^+/D^+\simeq 0.886$ turn out to be very similar and both comparable to the $p/\pi^+\simeq 0.745$, delineating the pure effect of canonical suppression owing to exact baryon-number conservation. This can be more quantitatively understood from the selective baryon-number canonical ensemble SHM, in which only baryon-number is subject to exact conservation in a canonical treatment but other quantum charges are still treated grand-canonically and thus conserved only {\it on average}~\cite{Vovchenko:2019kes}. This selective baryon-number canonical ensemble SHM can be obtained by putting all $\phi$'s other than $\phi_B$ to zero in the integrand of Eq.~(\ref{CE_partition_function}) and disregarding the corresponding integrals, yielding
\begin{align}\label{baryon-number-CE_partition_function}
Z(B)&=\frac{1}{2\pi}\int_0^{2\pi}d\phi_B e^{iB\phi_B}{\rm exp}[\sum_j \gamma_s^{N_{sj}} \gamma_c^{N_{cj}} e^{-iB_j\phi_B}z_j]\nonumber\\
&=\prod_{j=M^0,M^+,M^-}{\rm exp}(\gamma_s^{N_{sj}}\gamma_c^{N_{cj}}z_j)\times I_B(x_B),
\end{align}
where the contribution from mesons ($B_j=0$) is factorized out and the technique of separating the baryons ($B_j=+1$) and antibaryons ($B_j=-1$) leading to Eq.~(\ref{ZCE-mesons}) is used to arrive at the modified Bessel function of the first kind, with $x_B=\sum_{j=B,{\bar B}}\gamma_s^{N_{sj}}\gamma_c^{N_{cj}}z_j$ being the sum of multiplicities of all baryons and antibaryons in the grand-canonical limit. Substituting this into Eq.~(\ref{CE_mean_number}), the primary mean number of the $j$-th hadron from the baryon-number canonical ensemble SHM is
\begin{align}\label{selective-B-canonical-ensemble}
\langle N_j\rangle^{B-CE}=\gamma_s^{N_{sj}}\gamma_c^{N_{cj}}z_j\frac{I_{B-B_j}(x_B)}{I_B(x_B)}.
\end{align}
Therefore, with $x_B=2.38$ computed from the hadron spectrum we have employed at $V=20~{\rm fm^3}$, $T_H=170$\,MeV and $\gamma_s=\gamma_c=1$, the production of baryons and antibaryons in a completely neutral system is suppressed by a factor of $I_1(x_B)/I_0(x_B)\simeq 0.752$ as a result of the exact conservation of baryon number, which roughly explains the {\it relative} suppression of $p/\pi^+$, $\Lambda/K^0$ and $\Lambda_c^+/D^+$.

Similarly, when conserving only the charm number, the selective charm canonical ensemble SHM gives
\begin{align}
\langle N_j\rangle^{C-CE}=\gamma_s^{N_{sj}}\gamma_c^{N_{cj}}z_j\frac{I_{C-C_j}(x_C)}{I_C(x_C)},
\end{align}
where $x_C=\sum_{j, C_j=\pm1}\gamma_s^{N_{sj}}\gamma_c^{N_{cj}}z_j$ is the sum of multiplicities of all charm-hadrons in the grand-canonical limit.
With $x_C=0.01254$ in the present calculation, the pertinent chemical factor $I_1(x_C)/I_0(x_C)\simeq0.00627$ also roughly explains the {\it relative} suppression of $D_s^+/K^+\simeq 0.00599$, $\Lambda_c^+/p\simeq0.00688$ (where the common effect from canonical strangeness or baryon-number suppression is divided out) and $D^+/\pi^+\simeq0.00580$ (all computed from Table.~\ref{tab_CF-CEvsGCEmultiplicities}).

\subsection{System-size dependence of charm-hadron production}
\label{ssec_charm-system-size-dependence}
\begin{table*}[!t]
\begin{center}
\begin{tabular}{lcccccc}
\hline\noalign{\smallskip}
$CF$          & V=10~${\rm fm}^3$     & 20         & 50         & 100       & 200 \\
\noalign{\smallskip}\hline\noalign{\smallskip}
$D^0$         & 0.025877 &  0.066239  &  0.190294  &  0.373107 & 	0.627886 \\
$D^+$         & 0.025439 &  0.065891  &	 0.190002  &  0.372841 &	0.627669 \\
$D_s^+$       & 0.015805 &  0.053178  &	 0.178586  &  0.362376 &	0.619125\\
$\Lambda_c^+$ & 0.016956 &  0.055485  &  0.182039  &  0.365923 &	0.622147 \\
$\Xi_c^{+0}$  & 0.009884 &  0.042956  &	 0.167943  &  0.352535 &	0.611073 \\
$\Omega_c$    & 0.003495 &  0.022604  &	 0.130312  &  0.312514 &	0.576383 \\

\noalign{\smallskip}\hline\noalign{\smallskip}
$\Lambda_c^+/D^0$ & 0.655254  &	0.837649 & 	0.956620 &	0.980745 & 	0.990860 \\
$D_s^+/D^0$       & 0.610774  & 0.802820 &	0.938474 &	0.971239 &	0.986047 \\

\noalign{\smallskip}\hline
\end{tabular}
\end{center}
\caption{Chemical factors (CF) of ground-state charm-hadrons at $T_H=170$\,MeV, $\gamma_s=0.6$, $\gamma_c=15$ and varying volumes. The
ratios of the CF's of $\Lambda_c^+$ and $D^0$ and of $D_s^+$ and $D^0$ are also summarized in the last two rows.}
\label{tab_CF-vs-volumes}
\end{table*}

Having presented the analysis of the canonical suppression mechanisms, we are now in a position to study the system-size dependence of the charm-hadron yields
in a completely neutral system via CE-SHM under more realistic conditions. The strangeness saturation parameter is still fixed to be $\gamma_s=0.6$~\cite{He:2019tik}, as it has been shown to be insensitive to the system size for charged particle multiplicity $dN_{\rm ch}/d\eta\leq 50$~\cite{Vovchenko:2019kes} which covers the range of our current interest. For the charm fugacity factor $\gamma_c$, without further knowledge of the total charm cross section in the present $\sqrt{s}=13$\,TeV $pp$ collisions, we take a typical value of $\sim15$ determined from the charm canonical ensemble SHM in semi-central heavy-ion collisions at the LHC energy~\cite{Andronic:2006ky}. Given the flavor hierarchy in the operational hadronization temperature as suggested based on comparisons of light- and strange-quark susceptibilities~\cite{Bellwied:2013cta}, and the agreement of the susceptibilities computed from lattice with the same charm-baryon spectrum as used here up to temperatures of $170$\,MeV~\cite{Bazavov:2014yba}, we still take $T_H=170$\,MeV as our default value of hadronization temperature for the production of charm-hadrons~\cite{He:2019tik}. Uncertainty from using a possibly lower $T_H\sim 160$\,MeV may be (partially) bracketed by the varied branching ratios of the excited charm-baryons decaying into $\Lambda_c^+$~\cite{He:2019tik}.

In Table~\ref{tab_CF-vs-volumes}, chemical factors (CF) of ground-state charm-hadrons with $T_H=170$\,MeV, $\gamma_s=0.6$, $\gamma_c=15$ are displayed at varying volumes. At small volumes, $D_s^+$ and $\Lambda_c^+$ have smaller chemical factors than $D^0$ and $D^+$ due to additional canonical strangeness and baryon-number suppression, respectively. The chemical factors of charm-baryons $\Xi_c$ and $\Omega_c$ are further progressively smaller as the strangeness content increases, reflecting the canonical suppression from conserving all three (charm, strangeness and baryon number) quantum charges. As volume increases, the canonical suppression effects brought about by exact strangeness and baryon-number conservation are diminishing, and hence at asymptotically large volume, the chemical factors of all charm-harons tend to the same residual value that arises solely from their {\it common} canonical charm suppression. This is also seen from the gradual increase of the relative canonical suppression ({\it i.e.}, the ratios of the chemical factors) of $\Lambda_c^+/D^0$ and of $D_s^+/D^0$ toward unity from small to large volumes (see the last two rows of Table.~\ref{tab_CF-vs-volumes}). While the absolute value of the chemical factors does depend on $\gamma_c$, we have checked that the ratio of the chemical factors such as $\Lambda_c^+/D^0$ is nearly independent of $\gamma_c$, essentially guaranteeing that the fragmentation fraction of each charm-hadron is robust against variation of the $\gamma_c$. This can be understood from the fact that the $\Lambda_c^+/D^0$, with the {\it common} canonical charm suppression divided out (canonical suppression due to electric-charge conservation is negligible for very massive hadrons), can be attributed to the canonical baryon-number suppression on $\Lambda_c^+$ as discussed in Sec.~\ref{ssec_selectivecanonical} and thus intrinsically determined by the pertinent chemical factor $I_1(x_B)/I_0(x_B)$ in Eq.~(\ref{selective-B-canonical-ensemble}). But the charm-baryon contribution (proportional to $\gamma_c$) to the total $x_B$ turns out to be less than 2\% even for $\gamma_s=15$, so that the variation of $\gamma_c$ has little impact on the value of the relevant chemical factor.

\begin{table*}[!t]
\begin{center}
\begin{tabular}{lcccccc}
\hline\noalign{\smallskip}
$n_j~(\cdot 10^{-4}{\rm fm}^{-3}$)          & V=10~${\rm fm}^3$     & 20         & 50         & 100       & 200    & GCE\\
\noalign{\smallskip}\hline\noalign{\smallskip}
$D^0$         & 0.445553 &	1.148287  &	3.310131   &  6.495330 &	10.934662 	&	17.420949  \\
$D^+$         & 0.194705 &	0.503847  & 1.453016   &  2.851351 &	4.800262 	&	7.647869  \\
$D_s^+$       & 0.075040 & 	0.252484  &	0.847910   &  1.720531 &	2.939551 	&	4.747914\\
$\Lambda_c^+({\rm BR}50\%)$ & 0.126963 &	0.439135  &	1.497132   &  3.045487 &	5.207572 	&	8.415360 \\
$\Lambda_c^+({\rm BR}100\%)$& 0.149573 &	0.519555  &	1.776775   &  3.617118 &	6.187127 	&	10.001702 \\
$\Xi_c^{+0}$  & 0.016539 &	0.071955  &	0.281624   &  0.591389 & 	1.025276 	&	1.678110  \\
$\Omega_c$    & 0.000756 &	0.004889  &	0.028184   &  0.067592 &	0.124662 	&	0.216283 \\

\noalign{\smallskip}\hline\noalign{\smallskip}
$\Lambda_c^+/D^0({\rm BR}50\%)$ & 0.284956 	& 0.382426 	 & 0.452288 	& 0.468873 	&  0.476244 	&	0.483060  \\
$\Lambda_c^+/D^0({\rm BR}100\%)$& 0.335702 	& 0.452461 	 & 0.536769 	& 0.556880 	&  0.565827     &	0.574119 \\
$D_s^+/D^0$               & 0.168420    & 0.219879 	 & 0.256156     & 0.264887 	& 0.268829 	    &	0.272540  \\

\noalign{\smallskip}\hline
\end{tabular}
\end{center}
\caption{Total ({\it i.e.}, with feeddowns) thermal densities of ground-state charm-hadrons at $T_H=170$\,MeV, $\gamma_s=0.6$, $\gamma_c=15$ and varying volumes in the CE-SHM and also in the GCE-SHM (last column). The ratios of the thermal densities of $\Lambda_c^+$ and $D^0$ and of $D_s^+$ and $D^0$ are also summarized in the last three rows. For $\Lambda_c^+$, two scenarios of the branching ratios (BR=$50\%$ vs. BR=$100\%$) of the RQM-augmented excited $\Lambda_c$'s and $\Sigma_c$'s above $DN$ threshold decaying to the ground state $\Lambda_c^+$ are compared.}
\label{tab_densities-vs-volumes}
\end{table*}

We have computed the the primary production numbers of all charm-hadrons within the spectrum employed ({\it i.e.,} PDG charm-mesons, PDG charm-baryons, and additional charm-baryons from RQM predictions) with Eq.~(\ref{CE_mean_number}), and then the excited states are decayed into the ground state charm-hadrons with branching ratios specified in PDG. For excited $\Lambda_c$'s and $\Sigma_c$'s above the $DN$ threshold, we still assume two scenarios of BR=$100$\% or BR=$50$\% for their branching ratios decaying to the ground state $\Lambda_c^+$~\cite{He:2019tik}. In Table~\ref{tab_densities-vs-volumes}, the total ({\it i.e.}, with feeddowns from higher states) thermal densities (number divided by the corresponding volume) of ground-state charm-hadrons computed from the CE-SHM with $T_H=170$\,MeV, $\gamma_s=0.6$, $\gamma_c=15$ are summarized at varying volumes, in comparison with the GCE-SHM ones shown in the last column. An immediate observation is that the thermal density of each charm-hadron is increasing from small to large volumes toward their grand-canonical limit. The translated yield ratios of $\Lambda_c^+/D^0$ and $D_s^+/D^0$ also demonstrate a marked system-size dependence, representing a $\sim40$\% reduction from the largest to the smallest volume as shown here, essentially due to the additional canonical baryon-number and strangeness suppression on charm-baryons and charm-strange mesons in small systems. One notes that, while their thermal densities at $V=200~{\rm fm^3}$ are still far from the corresponding GCE-SHM values, the yield ratios of both $\Lambda_c^+/D^0$ and $D_s^+/D^0$ at this volume are already almost the same as the grand-canonical values, simply because the effects of canonical baryon-number and strangeness suppression already become vanishing and the residual canonical charm suppression is {\it common} for all charm-hadrons.

\section{Fragmenation and Decay Simulation
\label{sec_frag}}
To calculate the charm-hadron $p_T$ spectra and then their ratios, we take the charm quark $p_t$ spectrum generated by FONLL~\cite{Cacciari:1998it,Cacciari:2012ny} in $\sqrt{s}=13$\,TeV $pp$ collisions, and then perform its fragmentation into all kinds of charm-mesons and charm-baryons using the same procedure specified in~\cite{He:2019tik}. The fragmentation function~\cite{Braaten:1994bz} reads
\begin{align}
\label{frag}
D_{c\rightarrow H}(z)=N_H\frac{rz(1-z)^2}{[1-(1-r)z]^6}[6-18(1-2r)z
 \nonumber\\
+(21-74r+68r^2)z^2 \qquad \quad
\nonumber\\
-2(1-r)(6-19r+18r^2)z^3
\nonumber\\
\quad+3(1-r)^2(1-2r+2r^2)z^4]\,
\end{align}
where $z=p_T/p_t$ is the fraction of the hadron ($H$) momentum ($p_T$) with respect to the quark momentum ($p_t$). For the parameter $r$ in Eq.~(\ref{frag}), we take the same value $r_{D^0}$=0.1 and $r_{\Lambda_c^+}$=0.16 as used in~\cite{He:2019tik}, which were tuned to fit the slope of the measured $p_T$ spectra of the ground-state $D^0$ and $\Lambda_c^+$ in $\sqrt{s}=5.02$\,TeV $pp$ collisions. The value of $r$ for other charm-mesons ($M$) follows from the mass scaling: $r_M/r_{D^0}=((m_M-m_c)/m_M)/((m_{D^0}-m_c)/m_{D^0})$~\cite{Braaten:1994bz}, where $m_c=1.5$\,GeV is the charm-quark mass used in our calculations, and likewise for charm baryons ($B$): $r_B/r_{\Lambda_c^+}=((m_B-m_c)/m_B)/((m_{\Lambda_c^+}-m_c)/m_{\Lambda_c^+})$.

The normalization coefficient, $N_H$, for a given hadron, is determined by requiring its $p_T$ integrated yield to be equal to the fraction of its CE-SHM density at a given volume relative to the total density of all charm-hadrons; that is, fragmentation fractions are determined by the thermal weights as computed from CE-SHM. To associate the total mean number of the ground-state charm-hadrons $\langle N_j^{\rm tot}\rangle^{CE}$ computed in the CE-SHM with the rapidity density $dN_j/dy$ as experimentally measured in high-energy $pp$ collisions, one still need to establish the connection between the correlation volume $V_c$ entering the CE-SHM calculation in which quantum charges are strictly conserved and the $dV/dy$ corresponding to one unit of rapidity. Causal argument suggests that $V_c$ may correspond to a few units of rapidity~\cite{Castorina:2013mba}; canonical thermal fits also suggests that $V_c=(1$-$3)dV/dy$ and further $dV/dy\simeq 2.4dN_{\rm ch}/d\eta~({\rm fm^3})$~\cite{Vovchenko:2019kes}. In the following, $V_c=2dV/dy$ is chosen to demonstrate our final results, which has been found to yield best fit to the splitting of $\Lambda_c/D^0$ for the two bins of $dN_{\rm ch}/d\eta$ under consideration.

\begin{figure} [!tb]
\vspace{-0.5cm}
\includegraphics[width=1.05\columnwidth]{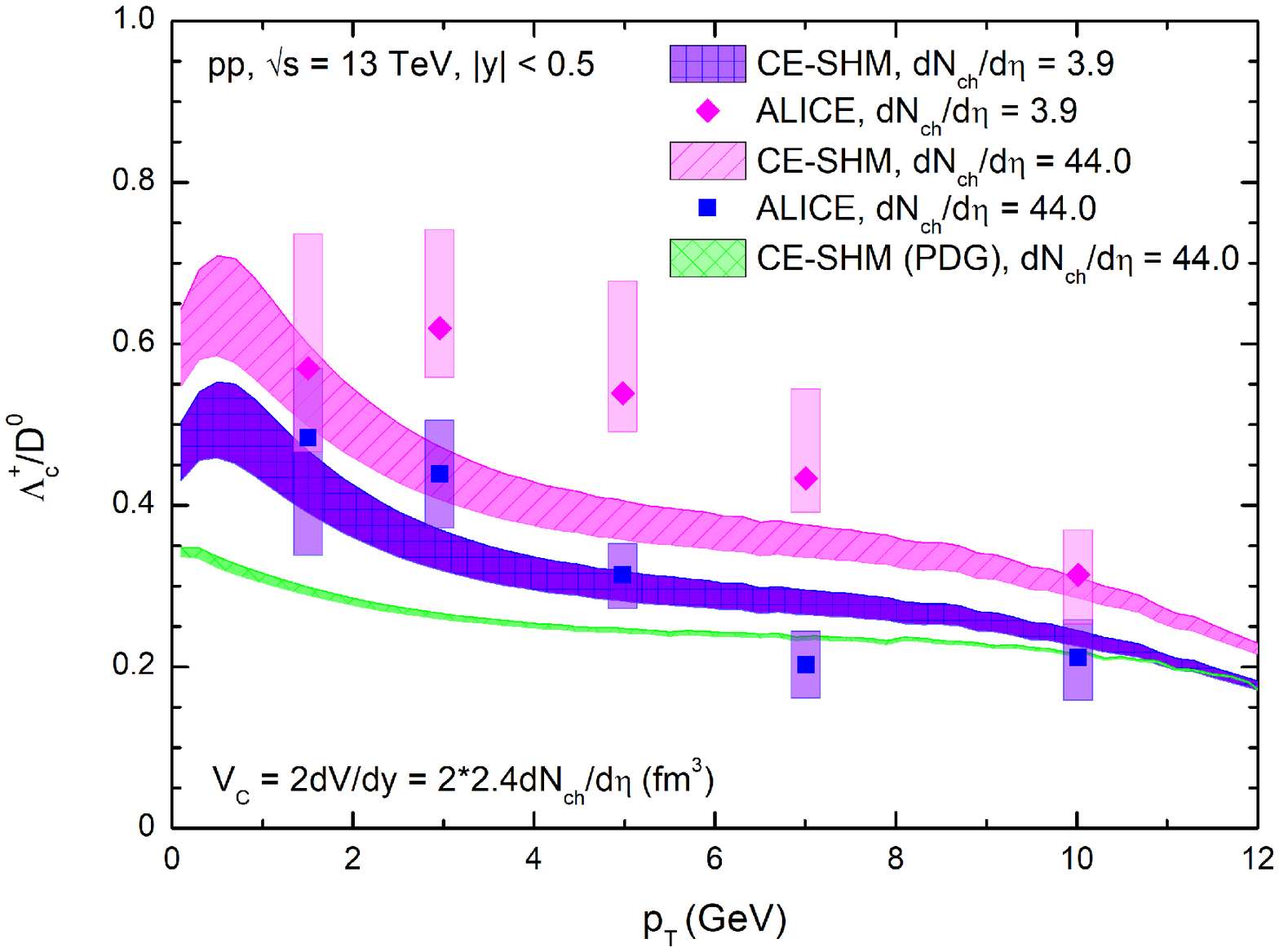}
\vspace{-0.5cm}
\includegraphics[width=1.05\columnwidth]{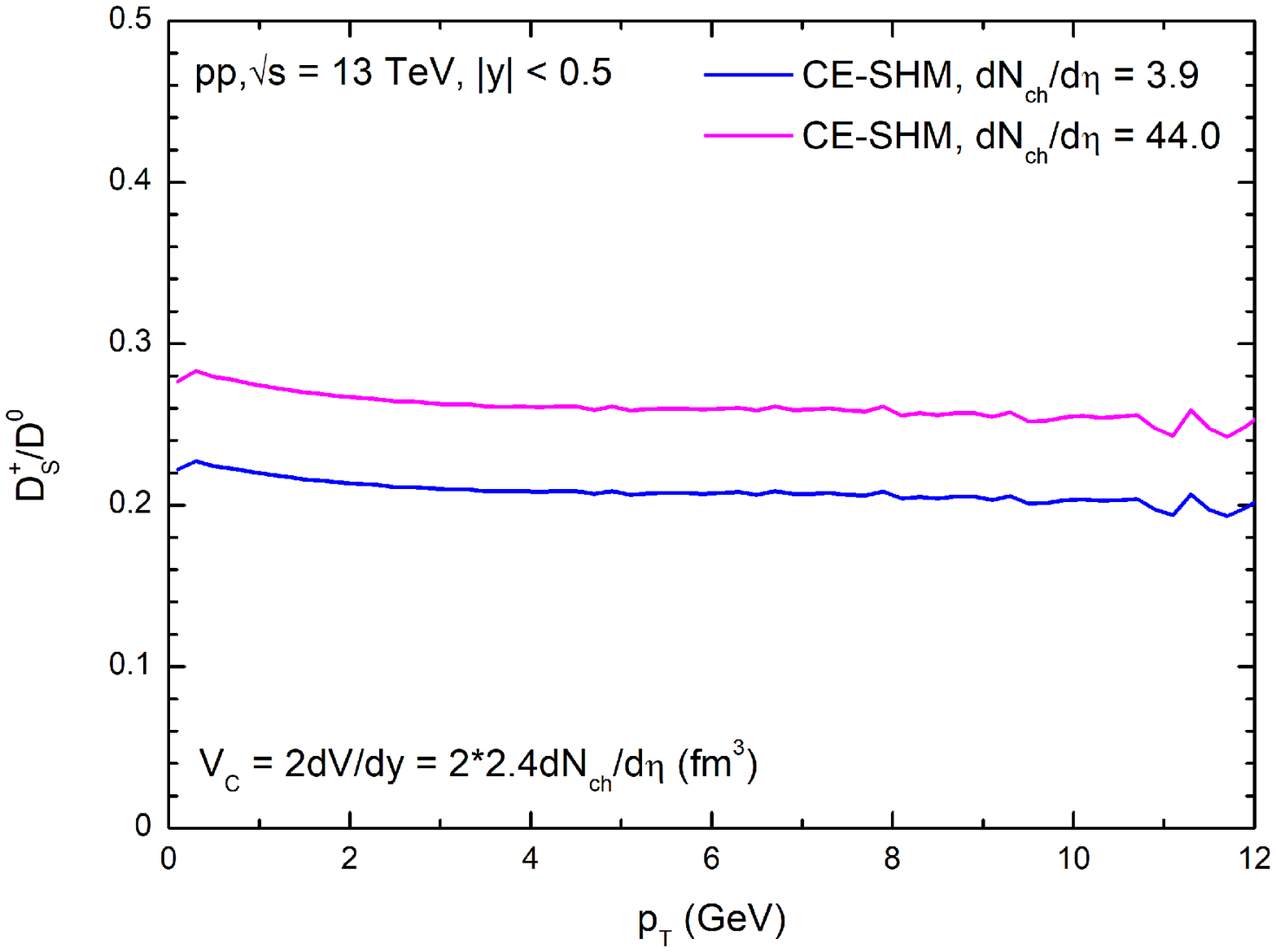}
\vspace{-0.5cm}
\caption{(Upper panel) The $\Lambda_c^+/D^0$ ratio computed from fragmentation (and decay) simulations with CE-SHM weights determined at $T_H=170$\,MeV, $\gamma_s=0.6$, $\gamma_c=15$ for two bins of charged particle multiplicity ($dN_{\rm ch}/d\eta=3.9$ and $44.0$) in $\sqrt{s}=13$\,TeV $pp$ collisions, in comparison with ALICE measurement~\cite{CHill:HP2020}. The correlation volume is taken to be $V_c=2dV/dy$ and its further relation with $dN_{\rm ch}/d\eta$ is quoted from~\cite{Vovchenko:2019kes}. The uncertainty band refers to the variation of the branching ratios (BR=$50$\%-$100$\%) of the RQM-augmented excited $\Lambda_c$'s and $\Sigma_c$'s decaying to the ground state $\Lambda_c^+$. For $dN_{\rm ch}/d\eta=44.0$, result from using only PDG listings of charm-baryons is also shown. (Lower panel) The same result for the $D_s^+/D^0$ ratio.}
\label{fig_Lc-Ds-over-D0}
\end{figure}

Each charm hadron produced from fragmentation is then decayed into ground-state particles by assuming a constant matrix element, with the decay kinematics solely determined by phase space and the branching ratios as discussed in Sec.~\ref{ssec_charm-system-size-dependence}. This way, the final total $p_T$-spectra (their respective absolute value associated with the total charm cross section is not our primary concern in the present study) of the ground state charm-hadrons are obtained and the $p_T$-dependent ratios of $\Lambda_c/D^0$ and $D_s^+/D^0$ are generated, as demonstrated in Fig.~\ref{fig_Lc-Ds-over-D0} with $V_c=2dV/dy$, with the uncertainty band in the $\Lambda_c/D^0$ encompassing the assumed range of the branching ratios (BR$=50$\%-$100$\%) of the RQM-augmented excited $\Lambda_c$'s and $\Sigma_c$'s decaying into the ground state $\Lambda_c^+$. We have also checked that varying $V_c=(2$-$3)dV/dy$ but fixing BR=100\% would generate almost the same central value and similar uncertainty band for the $\Lambda_c/D^0$.

From Fig.~\ref{fig_Lc-Ds-over-D0}, the role of the RQM-augmented charm-baryons is reaffirmed that generate a substantial enhancement of the $\Lambda_c/D^0$ relative to the case with only PDG listings of charm-baryons through their feeddowns to the ground state $\Lambda_c^+$~\cite{He:2019tik}. As the main result of the present study, the splitting of the $\Lambda_c/D^0$ between two bins of $dN_{\rm ch}/d\eta$ as indicated by the ALICE measurements is roughly described by our CE-SHM calculations, due to the additional canonical baryon-number suppression on charm-baryons that becomes more pronounced toward smaller system size. Yet the computed curve shows a significant deficiency relative to ALICE data at intermediate $p_T\simeq 2$-$5$\,GeV in the $dN_{\rm ch}/d\eta=44.0$ bin, which might be remedied by possible radial flow in such high multiplicity events hardening the charm-baryon spectrum more than the charm-meson one. Similarly, the calculated $D_s^+/D^0$ also exhibits a significant splitting between the two considered bins of $dN_{\rm ch}/d\eta$, which is attributed to the additional canonical strangeness suppression on charm-strange mesons.

\section{Summary
\label{sec_sum}}
We have computed the charged-particle multiplicity dependent hadro-chemistry of charm-hadrons (as represented by the $\Lambda_c/D^0$ and $D_s^+/D^0$ ratios) in high-energy $pp$ collisions within the approach of the canonical ensemble statistical hadronization model (CE-SHM). Taking full particle listings of PDG complemented by additional charm-baryons from relativistic quark model predictions, the canonical partition function has been evaluated and the chemical factors of typical hadrons that measure the pertinent canonical suppression computed. In particular, the mechanisms underlying the {\it relative} suppression between typical pairs of hadrons have been examined in connection with the selective canonical ensemble SHM that implements strict conservation of a single quantum charge (charm or baryon number) that distinguishes between the particles within the pair.

The chemical factors of various charm-hadrons have been shown to exhibit a pronounced system-size dependence that translates to charged-particle multiplicity dependence of charm-hadron ratios, highlighting the importance of strict conservation of quantum charges in small systems. Taking thus computed thermal densities of charm-hadrons as fragmentation weights, fragmentation and decay simulations yield $p_T$-dependent $\Lambda_c/D^0$ and $D_s/D^0$ ratios at different charged-particle multiplicities in fair agreement with ALICE measurements. The pronounced splitting of $\Lambda_c/D^0$ and $D_s/D^0$ between different multiplicity bins is attributed to the stronger canonical suppression of $\Lambda_c$ and $D_s$ in consequence of strict baryon-number and strangeness conservation (besides their common canonical charm suppression), respectively, toward smaller system-size.\\

{\bf Acknowledgments:}
This work was supported by the NSFC under grant 12075122. We are indebted to Ralf Rapp for useful communications.

\end{document}